\documentclass[conference]{IEEEtran}
\usepackage{cite}
\usepackage{amsmath,amssymb,amsfonts}
\usepackage{microtype}
\usepackage{subfigure}
\usepackage{wrapfig}
\usepackage{float}
\usepackage{array}
\usepackage{algorithmic}
\usepackage{graphicx}
\usepackage{textcomp}
\usepackage{xcolor}
\graphicspath{  {./images/}  }
\def\BibTeX{{\rm B\kern-.05em{\sc i\kern-.025em b}\kern-.08em
    T\kern-.1667em\lower.7ex\hbox{E}\kern-.125emX}}

\usepackage{breakurl}
\usepackage[breaklinks]{hyperref}
\begin{document}

\title{Decoding Imagined Auditory Pitch Phenomena with an Autoencoder Based Temporal Convolutional Architecture\\

}

\author{\IEEEauthorblockN{Sean Paulsen}
\IEEEauthorblockA{\textit{Department of Computer Science} \\
\textit{Dartmouth College}\\
Hanover, USA \\
e-mail: paulsen.sean@gmail.com}
\and
\IEEEauthorblockN{Lloyd May}
\IEEEauthorblockA{\textit{Computer Research in Music and Acoustics} \\
\textit{Stanford University}\\
Stanford, USA \\
e-mail: lloydmaydart@gmail.com}
\and
\IEEEauthorblockN{Michael Casey}
\IEEEauthorblockA{\textit{Department of Computer Science} \\
\textit{Dartmouth College}\\
Hanover, USA\\
e-mail: michael.a.casey@dartmouth.edu}

}

\maketitle

\begin{abstract}
Stimulus decoding of functional Magnetic Resonance Imaging (fMRI) data with machine learning models has provided new insights about neural representational spaces and task-related dynamics. However, the scarcity of labelled (task-related) fMRI data is a persistent obstacle, resulting in model-underfitting and poor generalization. In this work, we mitigated data poverty by extending a recent pattern-encoding strategy from the visual memory domain to our own domain of auditory pitch tasks, which to our knowledge had not been done. Specifically, extracting preliminary information about participants’ neural activation dynamics from the \textit{unlabelled} fMRI data resulted in improved downstream classifier performance when decoding heard and imagined pitch. Our results demonstrate the benefits of leveraging unlabelled fMRI data against data poverty for decoding pitch based tasks, and yields novel significant evidence for both separate and overlapping pathways of heard and imagined pitch processing, deepening our understanding of auditory cognitive neuroscience.
\end{abstract}

\begin{IEEEkeywords}
\textit{\textbf{neuroimaging; neuroscience; auditory cognition; deep learning.}}
\end{IEEEkeywords}

\section{Introduction}
\label{introduction}
\subsection{Motivation}
Brain decoding is the problem of classifying the stimulus that evoked given brain activity. Music's well-defined structure and the wealth of previous results about the neural representation of that structure are thus an appealing foundation upon which to approach this problem. Our primary goal was to train a machine learning classification model to predict the pitch-class of a note (the relative position of the note within the key) given an input of brain activity evoked by that note. We hypothesized that such a classifier would achieve significant results for three tasks: trained and tested on neural activity when the note is actually heard (hereafter referred to as the ``heard task"), the same when the note is only \textit{imagined} (``imagined task"), and most importantly, trained on neural activity when the notes are heard but evaluated on data when the notes are imagined (``cross-decoding task") to test for overlap between heard and imagined pathways. To our knowledge, the cross-decoding task had not been done before. Toward these ends, we obtained functional Magnetic Resonance Imaging (fMRI) data from musically trained participants while they both heard and imagined particular pitches. We further detail our scanning protocol in the Methods and Materials section. 

Training machine learning models on such voxel data is challenging, though, primarily due to the scarcity of relevant and labelled data to be used for training, and our experiments were no exception. However, Firat et al. \cite{Firat}'s work on visual memory brain decoding addressed this challenge of fMRI data poverty in a novel and effective way. More specifically, Firat et al. hypothesized that unlabelled fMRI data, which are normally deemed irrelevant and discarded, contain information about overall patterns of brain activity and can therefore be exploited in brain decoding classification tasks. Their architecture began with a sparse autoencoder \cite{Sparse} to perform unsupervised learning of neural activation patterns latent in unlabelled fMRI data. These patterns then served as filters in a temporal Convolutional Neural Network \cite{tCNN} to encode the labelled fMRI data into a non-linear, more expressive feature space. We refer to the inputs of this pipeline as ``unencoded datasets" and the outputs as ``encoded datasets" throughout this work. Thus, the encoded dataset is the result of filtering the task-dependent fMRI data by the patterns latent in task-independent data. Firat et al. then demonstrated improved performance of Multi-Voxel Pattern Analysis (MVPA) classifiers trained and tested on encoded datasets compared to unencoded datasets. 

\subsection{Our Approach} In Section 2 of this paper, we expand on the architecture of Firat et al. by adapting their autoencoder-tCNN pipeline from the visual domain to our novel auditory domain task of decoding imagined pitch. Section 3 presents our results, in which our encoded datasets are \textit{essential} for successful decoding of the imagined task, as well as first-of-their-kind significant results on the cross-decoding task. Section 4 discusses these results in the greater context of our goals and motivations. In particular, that this work demonstrates for the first time, to the best of our knowledge, that temporal filtering of fMRI data for an auditory task not only improves the performance of MVPA classifiers, but can also reveal fundamental, learnable attributes of auditory imagery that would go undetected by machine learning models trained on unencoded datasets. Section 5 details our methods and materials: participant selection, fMRI scanning protocol, hardware for training models, and statistical methods for evaluating our final classifiers. Section 6 concludes this paper and explores future work.
\begin{figure}

    \centering
    \includegraphics[width=0.4\textwidth]{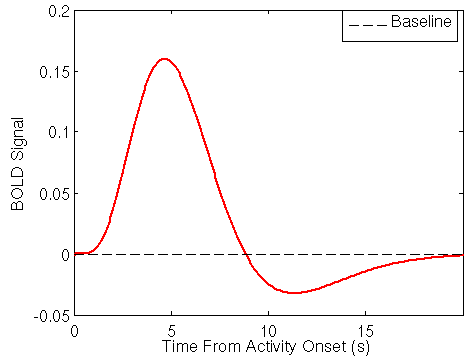}
    \caption{Hemodynamic Response Function (HRF) plotted as a 6-TR timeseries. \cite{HRF}.}
    \label{fig:hrff}
\end{figure}

\begin{figure*}[t!]

    \centering
    \includegraphics[width=0.6\textwidth]{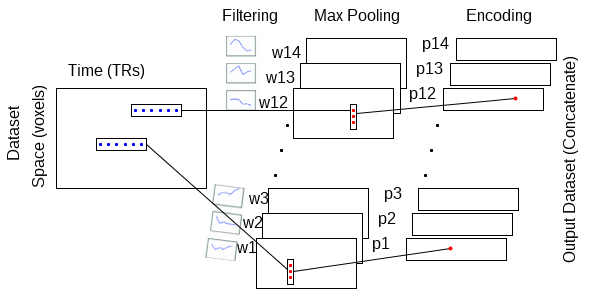}
    \caption{Our tCNN pipeline from voxel space to the encoded dataset. The filters are the neurons extracted from each trained autoencoder and represent neural activation patterns.}
    \label{fig:tcnn}
\end{figure*}

\section{Architecture Design}

\subsection{Neural Activation Pattern Training Data} Each fMRI scan yielded a timeseries of 3-dimensional voxel data, where the value of each voxel represented the intensity of neural activity at that geographic location in the brain. We used the Python Multi-Variate Pattern Analysis (PyMVPA) \cite{pyMVPA} library to store and transform fMRI data throughout the experiment. When we imported a participant's fMRI data, PyMVPA flattened the 3D voxel data into a single spatial dimension by concatenating along two axes (during which all voxels are preserved), restricted to one of twenty selected Regions Of Interest (ROIs) at a time, and provided a mapping back to 3D space for that ROI. Thus, we began with a matrix $VT$ of $V$-many voxels, which depended on each ROI, by $T$ timesteps, which was 1864 for all participants and ROIs. 

The Hemodynamic Response Function (HRF) in Figure \ref{fig:hrff} depicts the rise and fall of the intensity value of a voxel in response to a stimulus across 12 seconds. The time between images in our fMRI scans (TR) was 2 seconds, therefore the HRF would be observed across 6 timesteps in a given voxel. We thus expected any other latent activation patterns to occur across 6 timesteps as well. We therefore compiled our training data by sampling 1x6 windows of data from the matrix $VT$. Collecting every possible such window would provide the largest set of training data, but we believed the extreme overlap in that case could cause unpredictable bias during training. Spacing the samples out by exactly 6 timesteps would remove overlap, but could induce a different bias with every sample beginning and ending where another sample begins and ends, possibly limiting the kinds of patterns we expose to the model during training. Sampling with a stride greater than 6, however, might unnecessarily reduce the total size of our training set. Therefore our method considered each possible 6-TR window, then added it to the training data with probability 1/6. This allowed us to sample windows of training data that can begin at any timestep across the entire scan, while balancing our desire to both reduce overlap and minimize reduction of the training set. We further discarded any sample overlapping with labelled timesteps to avoid any possibility of downstream circularity. In summary, we collected 6-TR windows of unlabelled fMRI data, for each participant, for each ROI, to learn neural activation patterns latent in that participant in that ROI.

\subsection{Learning the Patterns} We implemented a sparse autoencoder model to perform unsupervised learning of the latent temporal neural activation patterns among each region's voxels without the need for hand-crafted features. The sparse autoencoder was implemented with the Keras \cite{keras} library in Python. The model input was encoded by a dense layer with sparsity enforced by an ``activity regularizer" parameter $\rho=.001$, hereafter referred to as the ``sparsity constraint," and then rectified linear unit (ReLU) activation functions were applied to obtain the encoded version of the input. We refer to the preceding steps as the ``encoding layer" throughout this paper. Each encoding layer had fourteen neurons in its dense layer, obtained via grid search on $\lbrace 8, 10, 12, 14, 16, 18\rbrace$. Each neuron's set of trained weights would then serve as a filter for obtaining the encoded dataset. The decoding layer was also dense, with six neurons (recall that this layer attempts to reproduce the six-dimensional input) and ReLU activations. The model was optimized via backpropagation to minimize the mean squared error between the output of the decoding layer and the input using the ``adamax" optimizer \cite{adam}.

\subsection{Filtering with Temporal Convolution} For each combination of participant and ROI, we extracted the set of learned neural activation patterns from the corresponding trained encoding layer and used them as filters in a tCNN to obtain the corresponding encoded dataset. Our tCNN pipeline is depicted in Figure \ref{fig:tcnn}. More specifically, we performed a 1D full convolution on the $VT$ matrix along its time axis with each of the fourteen trained neurons as the temporal filter. This resulted in fourteen response matrices for each combination of participant and ROI. Note that a full convolution means each response matrix had the same dimensions as $VT$. 

We expected the voxels to exhibit locally correlated activations \cite{pereira}, so we employed max pooling to extract spatial information from the filtered data in our response matrices. Recall, though, that $VT$ is the result of flattening the 3D voxel space to 1D, and therefore voxels next to each other in $VT$ are not necessarily next to each other geographically in the brain. Firat et al. \cite{Firat} did not detail their solution to this problem of 3D max-pooling with 1D data, so we devised our own method. Recall that PyMVPA provided a mapping back to the 3D voxel space of unencoded voxel values for each ROI, so we directly we backfilled the original 3D space with the values of each response matrix. 

For 3-dimensional spatial max-pooling, we proposed a pooling cube of tunable dimensions [$c_1$,$c_2$,$c_3$] moving exhaustively throughout each 3D space with no overlap, storing the maximum value within the cube at each step in a list. The jagged 3D voxel structure of each ROI was padded on all sides with zeroes due to the way PyMVPA maps back from 1D to 3D, so these zeroes needed to be accounted for. We certainly did not want to record a zero as a max-pooled value when the pooling cube is full of these padding zeroes, and more subtly we did not want to record voxel values on the jagged fringes as max-pooled values when they were being compared almost entirely to padding zeroes. Our solution was a tunable parameter $z_0$ which we called ``zero threshold''. The maximum value within the cube was only recorded as a max-pooled value when the proportion of non-zero values within the pooling cube exceeded $z_0$. Our [$2$, $2$, $2$] pooling cube and zero threshold of 0.6 were obtained via grid search.

We performed our method of 3D max-pooling on each timestep for each of the response matrices, applied hyperbolic tangent to each list of max-pooled values, and finally concatenated the lists for each timestep. The result of the concatenation was the encoded dataset for that participant and ROI. A repository of our code is available upon request.

\subsection{Pitch Decoding Classifiers} For each participant and ROI, we partitioned the labelled fMRI data by whether the corresponding pitch was heard or imagined. The heard samples were split further in half, with each half serving in turn as training data and testing data for an MVPA classifier. We stored the trained classifiers' predictions on the respective test sets with their corresponding pitch-class labels. Our analysis of classifier performance on the heard task was performed on the union of the two halves of test set predictions for each participant and ROI. The imagined task was evaluated similarly. For the cross-decoding task, we trained the classifier on all heard data, then predicted the labels of all imagined data. We calculated group level significance for each task and ROI using a t-test between per-participant prediction mean accuracies and null decoding model mean accuracies, detailed further in the Methods and Materials section. 

\section{Results}

\subsection{Temporal Filter Results} 
\begin{figure*}[t]

    \centering
    \includegraphics[width=0.5\textwidth]{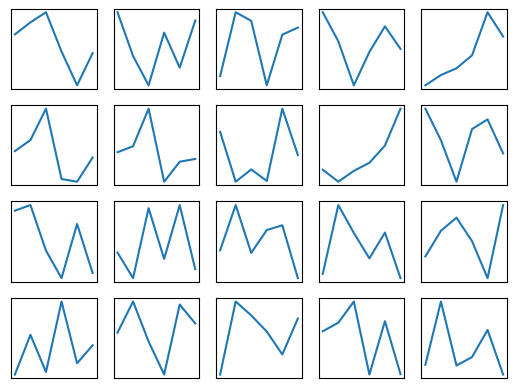}
    \caption{Learned temporal filters, sampled uniformly at random across all sparse autoencoders. Each consists of six weight values, one for each timestep. The HRF appears to have been learned by several of the selected neurons.}
    \label{fig:temporal filters}
\end{figure*}

Figure \ref{fig:temporal filters} shows twenty learned temporal filters (i.e, trained neurons) uniformly at random across the encoding layers of all participants and ROIs. Six weights connect each such neuron to the input layer, one for each timestep in the input, so we plotted the raw values of each sampled neuron's weights as a timeseries. This allows us to visually evaluate the learned filters as a pattern of neural activity. Observe that several of these patterns are good approximations of the HRF, which we expected most of the autoencoders to learn. Note further that none of the patterns are dominated by a single weight, which is to say that the models were not biased toward any particular timestep in the input data. This was the intent of our careful creation of each autoencoder's training data.

\subsection{Brain Decoding Results} Table \ref{table:pvalues} contains the results of our pitch decoding experiments. We evaluated the group-level statistical significance of the multivariate classifiers' ability to outperform chance in each of our regions of interest. The region of interest is given in the first column. The second column indicates the task, as explained above. The next two columns give the accuracy and False Discovery Rate (FDR)-corrected p-values when the classifiers were trained and evaluated with their respective encoded dataset, and the last two columns give the same information on the unencoded dataset. Observe one of our critical results, that thirteen of the fifteen successful regions \textit{required} the encoded dataset to obtain statistical significance. Eleven of the fifteen significant results were for the imagined task, and indeed \textit{all} of these regions required the encoded dataset for significance. 

\begin{table*}[t]
 \caption{Within-subject classifier results. FDR-corrected p-values for all ROIs with significant results. The encoded datasets enabled the classifiers to obtain significant results in thirteen of the fifteen significant regions.}
 \vskip 0.15in
\centering
\resizebox{\textwidth}{!}{%
    \begin{tabular}{||l|c|c|c|c|c||}
    \hline
    & & \multicolumn{2}{|c|}{\textbf{Encoded Dataset}} & \multicolumn{2}{|c||}{\textbf{Unencoded Dataset}}\\
    \hline
    \textbf{Region of Interest} & \textbf{Task} & \textbf{\begin{tabular}[c]{@{}l@{}}WPC Accuracy\\ Mean (Min, Max) \\ baseline =  0.1429\end{tabular}}  &\textbf{\begin{tabular}[c]{@{}l@{}}FDR-corrected pvals \\  (threshold = 0.05) \\(20 ROIs) \end{tabular}} & \textbf{\begin{tabular}[c]{@{}l@{}}WPC Accuracy\\ Mean (Min, Max) \\ baseline =  0.1429\end{tabular}}  &\textbf{\begin{tabular}[c]{@{}l@{}}FDR-corrected pvals \\  (threshold = 0.05) \\(20 ROIs) \end{tabular}}\\
    \hline
    Left Heschl’s Gyrus & H & 0.1642 (0.1250, 0.1964) & \textbf{0.0039}  &  0.1523  (0.0833, 0.2262)    &  0.5808   \\
    \hline
    Right Superior Temporal Sulcus & H &   0.1394  (0.0833, 0.2143)      &    0.8554     &    0.1754  (0.1190, 0.2560)     &  \textbf{0.0071}\\
    \hline
      Left Inferior Frontal Gyrus (Orbitalis)  &   I    &  0.1625 (0.0893, 0.2381)     &  \textbf{0.0368}     &  0.1485  (0.0952, 0.2024)     & 0.6475   \\
    \hline
        Left Precentral Gyrus  &    I     &   0.1607 (0.1190, 0.2262)     &     \textbf{0.0368}      &  0.1530  (0.0952, 0.2202)         &  0.5228   \\
    \hline
       Left Superior Temporal Gyrus   &   I      &  0.1684 (0.1190, 0.2202)     &    \textbf{0.0087}       &    0.1586  (0.1131, 0.2143)       &  0.2326   \\
    \hline
      Left Supramarginal Gyrus    &   I      &   0.1642 (0.0952, 0.2440)     &    \textbf{0.0355}       &    0.1502  (0.0893, 0.2083)       &  0.6291   \\
    \hline
       Left Insula   &   I      &   0.1649 (0.1310, 0.2440)    &     \textbf{0.0163}      &   0.1478  (0.0833, 0.2024)        &   0.6475  \\
    \hline
       Right Superior Temporal Sulcus  &    I     &   0.1604 (0.1131, 0.2083)    &     \textbf{0.0180}      &   0.1499  (0.0893, 0.2083)        &  0.6291   \\
    \hline
       Right Inferior Frontal Gyrus (Triangularis)   &   I      & 0.1688 (0.0952, 0.2500)      &   \textbf{0.0368}        &    0.1642  (0.1131, 0.2202)       &   0.0996  \\
    \hline
      Right Precentral Gyrus   &    I     &  0.1719 (0.1071, 0.2321)      &   \textbf{0.0103}        &    0.1569  (0.0952, 0.2560)       &  0.5228   \\
    \hline
       Right Superior Temporal Gyrus   &     I    &  0.1726 (0.1190, 0.2560)     &   \textbf{0.0124}        &   0.1453  (0.1012, 0.1845)        &  0.8149   \\
    \hline
       Right Supramarginal Gyrus   &     I    &   0.1656 (0.1250, 0.2440)    &    \textbf{0.0251}       &    0.1586  (0.0774, 0.2440)       &  0.5228   \\
    \hline
       Right Insula   &     I    &   0.1649 (0.1190, 0.2381)    &    \textbf{0.0124}       &    0.1506  (0.0893, 0.2024)      &  0.6291   \\
    \hline
       Right Superior Temporal Gyrus  &   X      &  0.1628 (0.1310, 0.1964)     &   \textbf{0.0157}        &   0.1492  (0.1071, 0.2083)        &   0.6445  \\
    
    \hline
        Right Rostral-Middle Frontal Gyrus & X &   0.1509 (0.1250, 0.1905)  &   0.3619     &    0.1642  (0.1131, 0.2143) & \textbf{0.0202} \\

    \end{tabular}%
    }
   
    \label{table:pvalues}

\end{table*}

\section{Discussion}

\subsection{Architecture Discussion}

Our first goal was to learn auditory neural activation patterns latent in 6-TR windows of unlabelled fMRI data with sparse autoencoders. We took care to avoid subtle biases when we collected our training data for the autoencoders by minimizing the overlap of the samples while allowing for the possibility of a sample to begin at any timestep in the scan. We plotted the weights of twenty uniformly randomly sampled encoder-layer neurons as timeseries to visualize the neural activation patterns that those neurons represented. These visualizations reassured our efforts in two ways. First, several of them are good approximations of the HRF, which we expected to be learned by one of the neurons in most of the autoencoders. Second, none of the patterns are dominated by a single timestep, and the peaks of activity are fairly well distributed across the timesteps, which was the intent of our training data collection method. These considerations, along with the success of our brain decoding classifiers, provide evidence that each neuron learned a latent auditory neural activation pattern, accomplishing our first goal. 

Our second goal was to generate a collection of encoded datasets by transforming the unencoded voxel data $VT$ in terms of the neural activation patterns learned by each autoencoder's encoding layer. Thus, the final step of our architecture was a modified tCNN. We used each of the learned activation patterns as temporal filters by convolving them with their respective $VT$ along the time axis and applying our own method of 3D max pooling. Concatenating the pooled matrices for each participant and ROI yielded our encoded datasets, thus achieving our second goal. Note that the final dimension after concatenating was dependent on the size of the 3D pooling cube and the number of filters. In our experiments the encoded datasets' dimensions ended up being roughly equal to the dimension of their respective unencoded dataset. However, one could increase the size of the pooling cube or learn fewer temporal filters if the dimensionality were a burden on computing. That would, of course, be a tradeoff with performance, but it is nevertheless valuable to have a mechanism for dimension reduction available in this pipeline.

\subsection{Brain Decoding Discussion}
Our third goal was to train a machine learning classifier to predict the pitch-class labels of heard and imagined pitches, trained and tested on fMRI data of twenty selected regions of interest. We hypothesized that such classifiers would outperform chance with statistical significance, and that the classifiers would achieve higher accuracy when trained on encoded datasets versus the unencoded datasets. We used the PyMVPA library to train multi-class Support Vector Machines (SVMs) with linear kernels on each of the encoded datasets and each of the unencoded datasets. Each classifier's accuracy was calculated on a held out test set, and the accuracies were averaged across participants for each ROI. Finally we calculated the group-level significance of the accuracies and controlled the FDR by correcting our p-values for multiple comparisons. Further details are in the Methods and Materials section. 

As shown in Table \ref{table:pvalues}, the statistical significance of outperforming chance relied almost entirely on the encoded datasets. For the imagined task the classifiers did not obtain significant results in any ROIs using \textit{unencoded} data. Indeed, training on the encoded datasets did not merely nudge almost-significant p-values past the threshold, but quite the opposite. Our encoded datasets enabled the classifiers to reduce their p-values by more than an order of magnitude in most regions in Table \ref{table:pvalues}, and \textit{two} orders in some, indicating that the encoded dataset reveals fundamental, learnable attributes of auditory imagery that would otherwise remain undetected by machine learning models trained on unencoded data. Thus, we achieved our third goal and obtained statistically significant evidence of our hypothesis in the case of the imagined task. Moreover, the significant results on the cross-decoding task provide a critical novel result- statistically significant evidence of geographical overlap between heard and imagined sound. 

Eleven of the fifteen significant results were achieved on the imagined pitch decoding task. This is explained by the greater cognitive involvement in imagining versus hearing sound. That is, imagining sound is a more involved activity than listening, evoking stronger, wider signals that are easier for the autoencoder to detect and learn. 

The heard and cross-decoding tasks both achieved two significant results, one each on the encoded and unencoded datasets. In both cases of significant unencoded datasets, the p-value for the respective encoded dataset was at least an order of magnitude worse. For the heard task, the two regions are near each other- Heschl's Gyrus and the Superior Temporal Sulcus are both auditory cortex areas in the superior temporal lobe. Therefore, while the inconsistency of the encoded dataset on the heard task requires further study, the results on the heard task are geographically consistent. On the other hand, the significant regions on the cross-decoding task are in separate lobes and non-adjacent. The Right Rostral-Middle Frontal Gyrus is interesting because significant results were achieved on the cross-decoding task with the unencoded dataset with a p-value at least an order of magnitude better than any other region for that task and dataset. Further, for the heard and imagined tasks, the encoded dataset improved the p-values in this region. Thus, the significant result in the Right Rostral-Middle Frontal Gyrus is curious, piquing further study.

\section{Methods and Materials}

\subsection{Participant Selection} Participants possessed at least 8 years of formal music training or professional performance experience in Western tonal music, and they completed the Bucknell Auditory Imagery Scale (BAIS) \cite{BAIS} and the Bregman Musical Ability Rating Survey \cite{BMARS}. Twenty-three such participants passed the screening process and provided their written informed consent in accordance with the Institutional Review Board at Dartmouth College. Each subject was compensated \$20 US upon completion of the scan.

All scanning used a 3.0 T Siemens MAGNETOM Prisma MRI scanner with a 32-channel head coil and Lumina button box with four colored push buttons. Each scan performed a T2* weighted single shot echoplanar (EPI) scanning sequence with a repetition time (TR) of 2 sec and 240mm field of view with 3mm voxels, yielding 80 voxel by 80 voxel images with 35 axial slices for a total of 224,000 voxels per volume. We used the fMRIPrep software \cite{fmriprep} to perform motion correction, field un-warping, normalization, and bias field correction preprocessing, as well as brain extraction and ROI parcellation, on the raw T2* BOLD data.

\subsection{fMRI Protocol} Each participant's fMRI scan consisted of 8 runs of 21 musical trials. Each scan was randomly assigned either the key of E Major or F Major, which was not known by the participant. We designed each run to collect data for either the heard task or the imagined task, alternating from run to run. Each trial began with an arpeggio in the assigned key for the participant to internally establish a tonal context, followed by a cue-sequence of ascending notes in their assigned major scale. After a randomized time interval, the participant either heard the next ascending note in the scale, or was instructed to imagine the next ascending note, depending on the run. The following four seconds (2 TRs) of scanning collected from all trials constituted the labelled data for the heard and imagined tasks. Next, a probe tone was played, and the participant rated the probe tone's goodness of fit in the tonal context from 1 to 4. We excluded the data of any participant with at least 20\% of their ratings missing, or whose ratings did not reflect internalization of the tonal hierarchy. Thus, we excluded the data of six of the twenty-three participants.   

Previous literature on imagined and heard tonal pitch-classes directed us to twenty regions of interest in the frontal, temporal, and parietal lobes according to the Desikan-Killiany (D-K) atlas in Freesurfer \cite{freesurfer}. The D-K ROIs are large cortical regions, reducing the burden of correcting for multiple comparisons compared to a larger quantity of smaller regions. Further, the D-K ROIs are consistent with the scales of relevant previous literature. The full table of the ROI atlas indices, cortical labels, and corresponding Brodmann areas is available on request. 

\subsection{Autoencoder Training}  The autoencoders were trained on Intel Xeon E5 processors, either 2.3, 2.6, or 3.2 GHz for 30 epochs on Dartmouth's Discovery High Performance Cluster with an average training time of approximately 3 hours. 10\% of the training data were held out as a validation set during training to prevent overfitting via early stopping. For each combination of participant and ROI, we trained ten autoencoders and kept the model with the lowest validation accuracy after 30 epochs. This was to avoid the rare but observed case where an autoencoder failed to find any minima during training.

\subsection{MVPA Classifiers} For each ROI, we partitioned the labelled fMRI data of each participant into two halves according to whether the pitches were heard or imagined. We then split the heard data in half, with each half serving in turn as training data and testing data for a multi-class SVM with linear kernels. We implemented the SVMs with the libSVM support vector machine library \cite{libSVM}. We then pooled the classifier's predictions on each of the two rounds of test data into a single set, along with their corresponding pitch-class labels. Our analysis of the heard task was performed on this collection of predictions and labels for each participant and region of interest. The imagined task was evaluated similarly. For the cross-decoding task, the classifier used all heard data for training, then predicted the labels of all imagined data. We calculated group level significance for each task using a t-test between per-participant prediction mean accuracies and null decoding model mean accuracies. We used Monte Carlo simulation to calculate the null models, repeating each classifier's training and testing 10,000 times with randomly permuted target labels and storing the mean overall accuracy. We corrected the group-level p-values for multiple comparisons using the method in Benjamini and Hochberg \cite{FDR}, which strictly controls the FDR of a family of hypothesis tests.

\section{Conclusion and Future Work}

In this work, we adapted the architecture and pipeline of Firat et al. \cite{Firat} from the visual domain to the auditory domain. Latent neural activation patterns were learned from unlabelled fMRI data, which are normally discarded, in order to generate our encoded datasets, which improved the performance of downstream MVPA classifiers. On the task of decoding the pitch class of imagined sound from fMRI data, the encoded datasets enabled the classifiers to outperform chance with group-level statistical significance in eleven ROIs. This demonstrated for the first time, to the best of our knowledge, that exploiting unlabelled fMRI data to perform temporal filtering for an auditory task not only improves the performance of MVPA classifiers, but can also reveal fundamental, learnable attributes of auditory imagery that would go undetected by machine learning models trained on unencoded datasets. Further, the group-level classifier performance on the cross-decoding task in two ROIs provided our novel statistically significant evidence of geographical overlap between heard and imagined sound. 

There are several immediate directions for future work. First is toward an end-to-end architecture for this task, rather than a disconnected training session to obtain the encoded datasets. Second is toward decoding/cross-decoding the other information in our fMRI protocol, such as the timbre (clarinet or trumpet) of the heard or imagined sound. Third is toward the generalization of our pipeline to other fMRI datasets with auditory tasks. Fourth is a deeper dive on the ROIs with significant cross-decoding results, as these results did not quite match our expectations.



\end{document}